\documentclass[preprint,tightenlines,aps,floats]{revtex4}

\usepackage{epsf}
\usepackage{epsfig}
\usepackage{graphicx}
\usepackage[dvips]{color}
\usepackage{amsmath,amssymb,bbm,mathrsfs}

\newcommand{\tpi}{\tilde{\pi}}
\newcommand{\trho}{\tilde{\rho}}

\newcommand{\notD}{\hbox{{$D$}\kern-.60em\hbox{/}}}
\newcommand{\notE}{\ \hbox{{$E$}\kern-.60em\hbox{/}}}
\newcommand{\notp}{\ \hbox{{$p$}\kern-.43em\hbox{/}}}

\newcommand{\Del}{\Delta}

\includegraphics{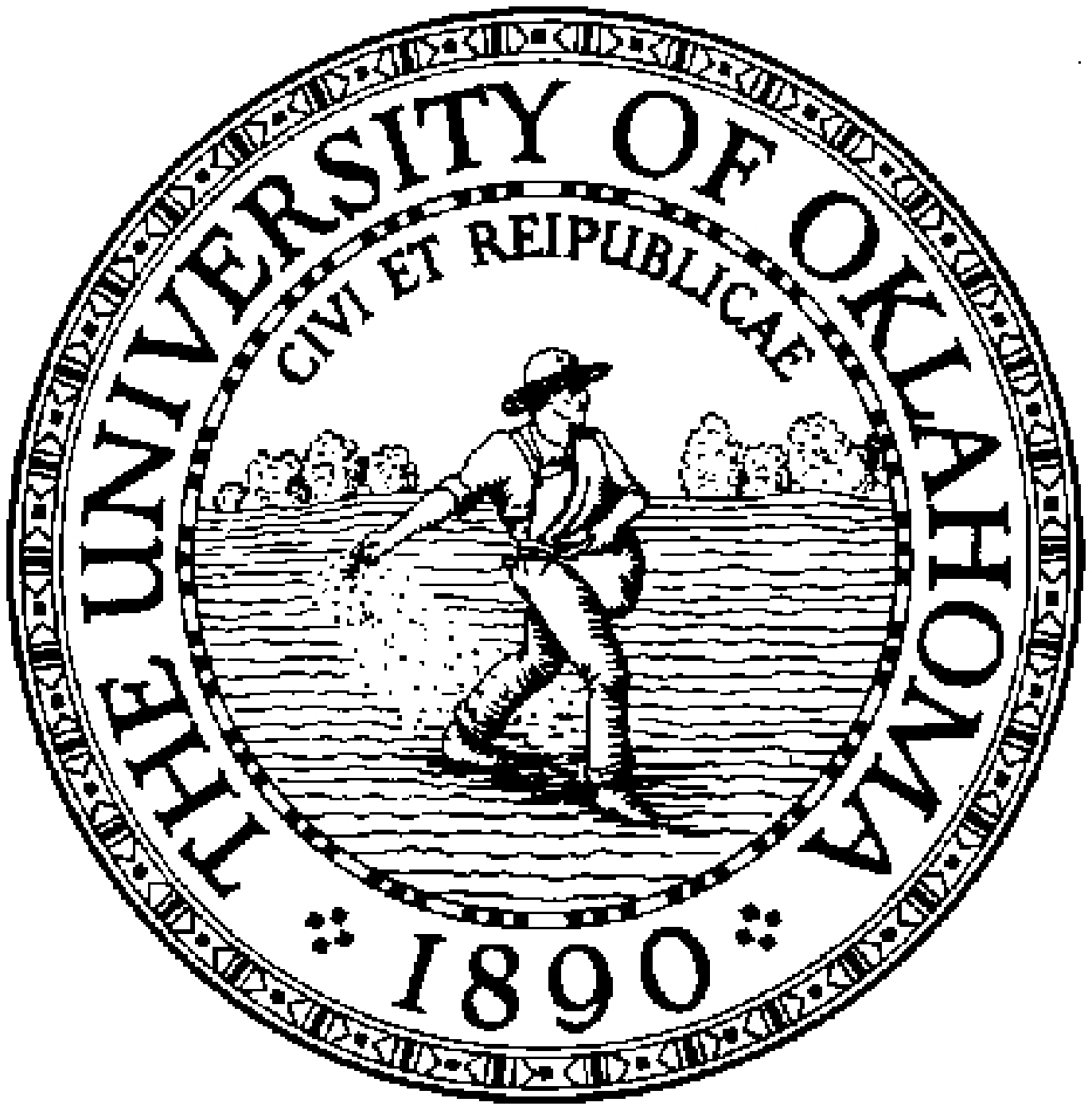}

\preprint{\font\fortssbx=cmssbx10 scaled \magstep2
\hbox to \hsize{
\hskip1.2in 
\hbox{\fortssbx The University of Oklahoma}
\hskip0.2in $\vcenter{
                      \hbox{\bf arXiv: [hep-ph]}
                      \hbox{\bf OSU-HEP-10-11}
                      \hbox{\bf OU-HEP-101222}
                      \hbox{December 2010}}$ }
}

\begin{document}

\title{\vspace*{0.7in}
Discovering Colorons at the Early Stage LHC}

\author{
Duane A. Dicus$^a$\footnote{Email address:dicus@physics.utexas.edu},
Chung Kao$^b$\footnote{Email address: kao@physics.ou.edu},
S. Nandi$^c$\footnote{Email address: s.nandi@okstate.edu},
Joshua Sayre$^b$\footnote{Email address: sayre@physics.ou.edu}}

\affiliation{
$^a$Center for Particles and Fields and Texas Cosmology Center, 
University of Texas, Austin, TX 78712, USA \\
$^b$Homer L. Dodge Department of Physics and Astronomy
and Oklahoma Center for High Energy Physics,
University of Oklahoma,
Norman, OK 73019, USA \\
$^c$Department of Physics and Oklahoma Center for High Energy Physics,
Oklahoma State University, Stillwater, OK 74078, USA
\vspace*{.4in}}

\date{\today}

\begin{abstract}

We investigate the prospects for the discovery of massive hyper-gluons
using data from the early runs of the CERN Large Hadron Collider with 
$\sqrt{s} = 7$ TeV and assuming an integrated luminosity of 1 fb$^{-1}$.
A phenomenological Lagrangian is adopted to evaluate the cross section
of a pair of colored vector bosons (coloron, $\tilde{\rho}$) decaying into
four colored scalar resonances (hyper-pion, $\tilde{\pi}$),
which then decay into eight gluons.
We include the dominant physics background from the production of 
$8g$, $7g1q$, $6g2q$, and $5g3q$.
We find an abundance of signal events and that realistic cuts 
reduce the background enough to establish a $5\sigma$ signal for 
$m_{\tilde{\pi}} \alt 220$ GeV or $m_{\tilde{\rho}} \alt 733$ GeV.

\end{abstract}

%


\maketitle


{\sl Introduction.}--
With the LHC beginning to accumulate data, we look forward to a new
era of high-energy physics as we explore multi-TeV energy scales.
In addition to the search for the Higgs boson as a completion of the
Standard Model, many scenarios have been considered for the discovery
of new physics.
Often, the discovery potential provided by the
LHC's unprecedented collision energies is mitigated by the prevalence
of jets derived from Standard Model processes. New physics which
proceeds through weak interactions, such as Higgs production, must be
carefully separated from large, strong-force produced backgrounds via
judicious selection cuts when jets are involved.

It is also possible that new physics will manifest itself
through the strong force. If new colored particles exist at TeV
scales, they will be discovered through decays into jets. One generic
possibility is a massive vector boson in the color-octet
representation~\cite{Hill:1991at,Dicus:1994sw,Dobrescu:2007yp}.
Such a particle has been dubbed a coloron.
According to this scenario
several theories of physics beyond the Standard
Model give rise to colorons such as
topcolor model~\cite{Hill:1991at} and Kaluza-Klein excitations of
the gluon in  the universal extra-dimensional models~\cite{
Appelquist:2000nn,Dicus:2000hm}.
In Ref.~\cite{Kilic:2008pm}, Kilic, Sundrum and
Okui showed how a coloron, as well as a scalar octet, can emerge as
the low energy states of an effective theory arising from a simple
model of new, strongly interacting fermions charged under this 
new strong interaction as well as QCD color. 
In this letter we follow
their analysis closely, as well as the subsequent treatment found in
Ref.~\cite{Kilic:2008ub}.

Briefly, if we suppose that there exists a new, strong 
force, termed hypercolor, it may become confining at higher energies
compared to the strong QCD force. Fermions which carry hypercolor will
form bound states which are hypercolor singlets but which may carry
QCD color quantum numbers. In particular, if these ``hyperquarks" are also
triplets of QCD then their lightest bound states will be color octets. 
These are like the $\rho$ meson octet of ordinary  QCD, but are
hypercolor singlet and color octet. 
Analogous to the breaking of chiral symmetry in the standard
model, this model will produce relatively light hyper-pions 
as pseudo-Goldstone bosons in the octet representation.

These colorons, and scalar octets can have an interesting phenomenology.
Naively one might think that light octets are severely constrained by
dijet searches at the Tevatron. However, in this model
the hyper-pions couple sufficiently weakly to gluons to leave an
interesting parameter range and the hyper-rhos have only a small branching
fraction to decay into two quarks or two gluons. 
Rather, the hyper-rho decays predominantly to hyper-pions, 
each of which then decays to a gluon pair. 
Thus the dominant signal for resonant production of each
hyper-rho is a 4-jet decay chain.

On the other hand, there are several processes which pair produce
hyper-pions without a resonant hyper-rho. Combined with the loss of jet
resolution during showering, hadronization, and reconstruction, this
may make the initial hyper-rho resonance difficult to establish. 
To better find it we consider the pair production of hyper-rhos,
leading to an eight-jet signal and consider this signal as a
potential discovery in the near future. 
Before refurbishing, the LHC is running at a 
center of mass energy of 7 TeV.
We present results for possible detection at this energy
assuming an integrated luminosity of $1$ fb$^{-1}$.

\medskip

{\sl A Model with Colored Vector Bosons and Scalars.}--
  As detailed in Ref.~\cite{Dicus:1994sw,Kilic:2008pm,Kilic:2008ub},
  we assume there is a new
  $SU(N_{\rm HC})$ gauge group, hypercolor, acting on a new set of fermions
  which also carry Standard Model color charges. By analogy with the
  spontaneous breaking of chiral symmetry in the Standard Model, one
  can derive an effective Lagrangian for new, massive color
  octets. These would be visible as a set of scalars, designated
  $\tilde{\pi}$, and a vector boson, $\tilde{\rho}$. Their
  interactions are expressed by the following effective
  Lagrangian~\cite{Kilic:2008ub}:
%
%
\begin{eqnarray}
{\cal L}_{\rm eff}
& = & -\frac{1}{4}G_{\mu\nu}^a G^{a\mu\nu} +\bar{q}i\notD q
 -\frac{1}{4}\trho_{\mu\nu}^a\trho^{a\mu\nu}
 +\frac{M_{\trho}^2}{2}\trho_\mu^a\trho^{a\mu}
 -g_3\epsilon \trho^a_\mu\bar{q}\gamma^\mu T^a q  \nonumber\\
&  & +\frac{1}{2}(D_\mu \tpi)^a (D^\mu \tpi)^a -{M_{\tpi}}^2\tpi^a\tpi^a
 -i g_{\trho\tpi\tpi}f^{abc}\trho^{a\mu}(\tpi^b D_\mu\tpi^c) \nonumber\\
&  & -\frac{3{g_3}^2 \epsilon^{\mu\nu\rho\sigma}}{16\pi^2 f_{\tpi}}
\, {\rm Tr}[\tpi G_{\mu\nu}G_{\rho\sigma}]
 +i\chi g_3\, {\rm Tr}[G_{\mu\nu}[\trho^\mu,\trho^\nu]] \,.
%
\end{eqnarray}
The number of hypercolors has been set to be $N_{\rm HC} = 3$ for simplicity.
$G_{\mu,\nu}$ and $q$ are Standard Model (SM) gluon and
  quark fields, while $a$ is a color index. Under the assumptions of
  the model, Kilic et al. have derived most of the parameters in terms
  of a single variable, $m_{\tilde{\rho}}$. 
They set $\epsilon \simeq 0.2$, $g_{\trho\tpi\tpi} \simeq 6$, 
$M_{\tpi} \simeq 0.3\times M_{\trho}$, and
\begin{equation}
 f_{\tpi} \simeq  f_{\pi}\times \frac{M_{\trho}}{m_{\rho}} \, .
\end{equation}
We use exactly these relations so a value of $m_{\tilde{\pi}}$
determines the value of $M_{\tilde{\rho}}$.
The last term of the Lagrangian in Eq.~(1) contains a free parameter
$\chi$ which cannot be extrapolated from the Standard Model.

\medskip

{\sl Production Cross Section.}--
In our model the colorons, $\trho$, can only be pair produced  
at the LHC via gluon fusion.
(Single production of a coloron resonance is suppressed due to the
small $q \bar{q}\trho$ coupling,
as well as small quark-antiquark luminosity at the LHC). 
We calculate the cross section at the early LHC for
$pp \to \trho\trho \to 4\tpi \to 8g +X$
with the parton distribution functions of CTEQ6L1 \cite{CTEQ6}.
The factorization scale as well as the renormalization scale is chosen
to be (a) the coloron mass ($M_{\trho}$) for the coloron signal and
(b) the root mean square transverse momentum
($\sqrt{\langle p_T^2 \rangle}$)
of all eight jets for the physics background,
with the leading order evolution of the strong coupling.
For simplicity, the $K$ factor is taken to be one for both the signal
and the background.

We have evaluated the cross section for
$pp \to \trho\trho \to 4\tpi \to 8g +X$ from gluon fusion
and quark-antiquark fusion in two ways, 
(a) with complete matrix element involving Breit-Wigner resonances of
both $\trho$ and $\tpi$, and
(b) with matrix elements involving Breit-Wigner resonance of $\trho$ for
$pp \to \trho\trho \to 4\tpi +X$ and narrow width approximation (NWA)
for $\tpi \to gg$.
A new model has been added in MadGraph~\cite{Stelzer:1994tk,Murayama:1992gi}
with new interactions and new particles to generate matrix elements 
squared for all processes in (a) and (b).
In the narrow width approximation, the cross section for
$pp \to \trho \trho \to 4\tpi \to 8g +X$ can be thought of as
the production cross section $\sigma(pp \to \trho\trho \to 4\tpi +X)$
multiplied by the branching fraction of hyper-pions decay into gluon pairs
$B(\tpi \to gg) = 1$.
In addition, we have checked
$|M|^2(gg \to \trho\trho \to 4\tpi)$ analytically.
The numerical output from MadGraph gives excellent agreement
with that from our analytic expressions.

With energy-momentum smearing, the cross section in
the narrow width approximation (NWA)
agrees very well with that evaluated via a Breit-Wigner resonance
(BWR) for most parameters that we have chosen.
The ATLAS detector specifications~\cite{ATLAS} have been adopted
to model these effects by Gaussian smearing the momenta of the jets, 
\begin{eqnarray}
\frac{\Delta E}{E} = \frac{0.60}{\sqrt{E}} \oplus 0.03 \, ,
\end{eqnarray}
with individual terms added in quadrature.


\begin{figure}[htb]
\centering\leavevmode
\epsfxsize=4.4in\epsffile{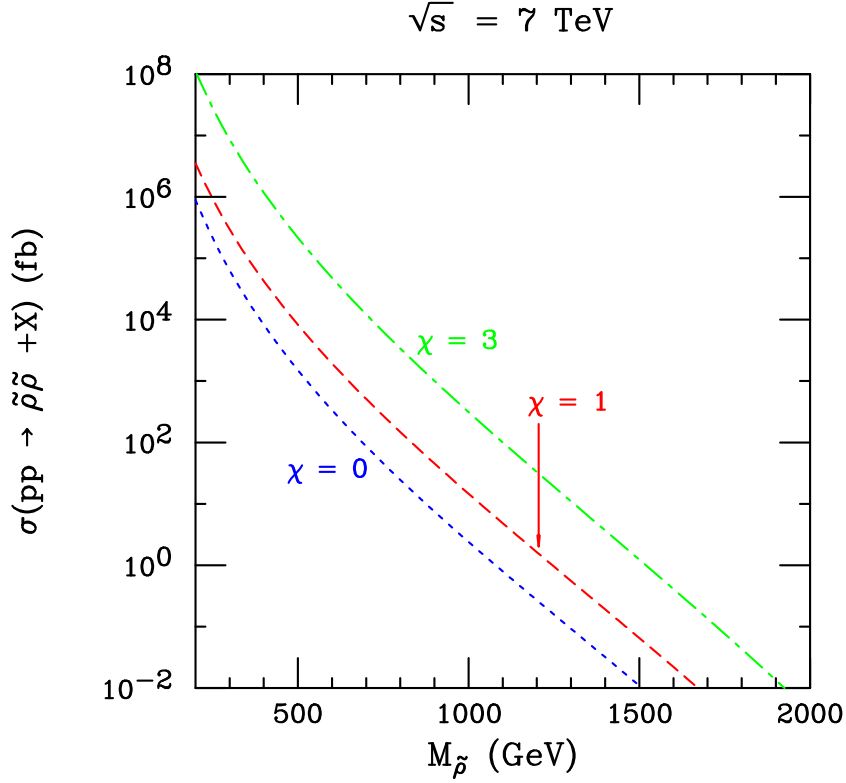}

\caption[]{
The cross section of $pp \to \trho\trho +X$ for $\chi = 1, 0$ and $3$
at the LHC with $\sqrt{s} = 7$ TeV, as a function of $M_{\trho}$.
\label{fig:sigma1}
}\end{figure}

To demonstrate that colorons can be produced copiously at the early
LHC, we show, in Fig. 1, the cross section for
$pp\rightarrow\,\tilde{\rho}\tilde{\rho}\,+\,X$
for a few values of $\chi$.
For this figure we took the scale to be $Q = M_{\trho}$, and did not
make any cuts.
For $gg\,\rightarrow\,\tilde{\rho}\tilde{\rho}$
we use an analytic expression for the square of the matrix element,
summed over polarizations,
\begin{eqnarray}
\sum_{{\rm pol}}|T|^2\,
 &=&\,\frac{Y^2(1-z^2)^2}{(1-\beta^2z^2)^2}\frac{E^4}{M_{\trho}^4}
     \left[12-12Y+(5+z^2)Y^2\right]  \nonumber \\
 &+&\,\frac{Y^2(1-z^2)}{(1-\beta^2z^2)^2}\frac{E^2}{M_{\trho}^2}
     \left[16(1+3z^2)-2(11+18z^2)Y+(5+9z^2+3z^4)Y^2\right]  \nonumber \\
 &+&\,\frac{1}{(1-\beta^2z^2)^2}\bigg[8\left(16+3\frac{M_{\trho}^4}{E^4}\right)
      -256Y+(160+16z^2+36z^4)Y^2  \nonumber  \\
 &&\,\,\,\,\,\,\,\,\,\,\,\,\,\,\,\,\,\,\,\,\,\,\,\,\,\,\,\,\,\,\,\,\,\,\,\,\,\,\,\,\,\,\,\,\,\,-(32+22z^2+24z^4)Y^3+(2+5z^2+4z^4+2z^6)Y^4\bigg]  \nonumber  \\
 &+&\frac{1}{1-\beta^2z^2}\left[-6\left(16+4\frac{M_{\trho}^2}{E^2}
    +\frac{M_{\trho}^4}{E^4}\right)+140Y
    -(58+24z^2)Y^2+3(1+4z^2)Y^3-z^4Y^4\right]  \nonumber \\
 &+&\,28+6\frac{M_{\trho}^2}{E^2}-3(1-\beta^2z^2)-16Y+4Y^2
\end{eqnarray}
where $E$ is the gluon energy,
$z$ is the cosine of the scattering angle, 
$\beta^2\,=\,1-M_{\trho}^2/E^2$,
and $Y\,=\,1-\chi$.
Clearly the theory is only unitary for $\chi\,=\,1$ where the terms
which grow with energy are absent;
in the remainder of the paper that is the only value we will use.
For $\chi\,=\,1$ we have checked that our results
for $gg \to \trho\trho$ are consistent with
those of Refs.~\cite{Dicus:1994sw,Dobrescu:2007yp,Kilic:2008ub}
for $\sqrt{s}\,=\,14$ TeV.

\medskip

{\sl Physics Background.}--
We compute the cross section of the dominant eight jet physics background
with the matrix-element generator COMIX~\cite{Gleisberg:2008fv},
interfaced with the event generator SHERPA~\cite{Gleisberg:2003xi}.
COMIX adopts color-dressed Berends-Giele
recursion relations~\cite{Berends:1987me,Duhr:2006iq}
to construct QCD amplitudes. 
The physics backgrounds included are, in order of importance,
$gq \to 7g1q$, $gg \to 8g$, $qq \to 6g2q$, and $gq \to 5g2q1\bar{q}$. 
For $M_{\tpi} \agt 230$ GeV,
the $6g2q$ process becomes larger than the $8g$ process.

MadGraph employs traditional Feynman diagrams. It can calculate
matrix elements with at most five outgoing gluons from gluon fusion. 
We have compared the cross
section of $gg \to 4g$ with MadGraph and COMIX and have found
excellent agreement.

To reduce the large QCD physics background, we require that in each event
there should be eight jets ($j = g,q,\bar{q}$) with lower limits on
their transverse momenta of 
\begin{equation}
p_T(j_1, \cdots, j_8) \ge 250,200,160,120,80,60,40,20 \; {\rm GeV} \, ,
\end{equation}
a pseudo-rapidity for each jet of $|\eta(j)| < 2.5$, 
and angular separation for each pair of jets
$\Delta R = \sqrt{\Delta\phi^2 +\Delta\eta^2} > 0.5$.

\medskip

{\sl Discovery Potential at the Early LHC.}--
To study the discovery potential of
$pp \to \trho\trho \to \tpi\tpi \to 8g +X$, 
we evaluate cross sections for the SM backgrounds as described 
in the previous section. In addition, we have considered two types
of mass cuts: (i) relative mass cuts and (ii) fixed mass cuts.

The relative mass cut requires that within each event there must be
eight jets, which can be arranged into four pairs of jets
that have invariant mass within $\Delta M_{2j}$ of one another, 
and there must be distinct pairs of four jets that have invariant mass
within $\Delta M_{4j}$ of each other.
We have chosen
(a) $\Delta M_{2j} \le 30$ GeV and $\Delta M_{4j} \le 60$ GeV
or
(b) $\Delta M_{2j} \le 50$ GeV and $\Delta M_{4j} \le 100$ GeV.

The fixed mass cut requires that within each event, there must be
eight jets with four pairs of jets that have invariant mass
within a bin of width $\pm\Delta M_{2j}$ centered at $M_{\tpi}$, 
and there must be two groups of four jets that have invariant mass
within a bin of width $\pm\Delta M_{4j}$ centered at $M_{\trho}$.
We have chosen
(a) $|M_{2j}-M_{\tpi}|  \le 0.10M_{\tpi}$ and
    $|M_{4j}-M_{\trho}| \le 0.15M_{\trho}$
or
(b) $|M_{2j}-M_{\tpi}|  \le 0.15M_{\tpi}$ and
    $|M_{4j}-M_{\trho}| \le 0.20M_{\trho}$.

We define the signal to be observable
if the lower limit on the signal plus background is larger than
the corresponding upper limit on the background \cite{HGG}, namely,
\begin{eqnarray}
L (\sigma_s+\sigma_b) - N\sqrt{ L(\sigma_s+\sigma_b) } >
L \sigma_b +N \sqrt{ L\sigma_b }
\end{eqnarray}
which corresponds to
\begin{eqnarray}
\sigma_s > \frac{N^2}{L} \left[ 1+2\sqrt{L\sigma_b}/N \right] \, .
\end{eqnarray}
Here $L$ is the integrated luminosity assumed to be 1 fb$^{-1}$,
$\sigma_s$ is the cross section of the coloron signal,
and $\sigma_b$ is the background cross section.
The parameter $N$ specifies the level or probability of discovery.
We take $N = 2.5$, which corresponds to a 5$\sigma$ signal.

To assess the discovery potential we present in Fig. 2 
the cross sections of the coloron signal, and the physics background,  
after acceptance cuts and relative mass cuts, versus $M_{\tpi}$. 
Also shown are the background cross section for the SM processes
with the relative mass cuts discussed above.
In addition, we present the minimal signal cross section that is required
to establish a $5\sigma$ signal with relative mass cuts (a) or (b)
as given above.

We note that a narrower relative mass cut (a) has the potential to
discover the colorons and hyper-pions up to $M_{\tpi} = 200$ GeV 
($M_{\trho} = 667$ GeV). A wider relative mass cut (b) will allow more
background events, and thus has a slightly reduced discovery reach of
$M_{\tpi} = 180$ GeV ($M_{\trho} = 600$ GeV). 
In addition, relative mass cut (a) can improve the signal to
background ratio ($\sigma_s/\sigma_b$) significantly.


\begin{figure}[htb]
\centering\leavevmode
\epsfxsize=4.4in\epsffile{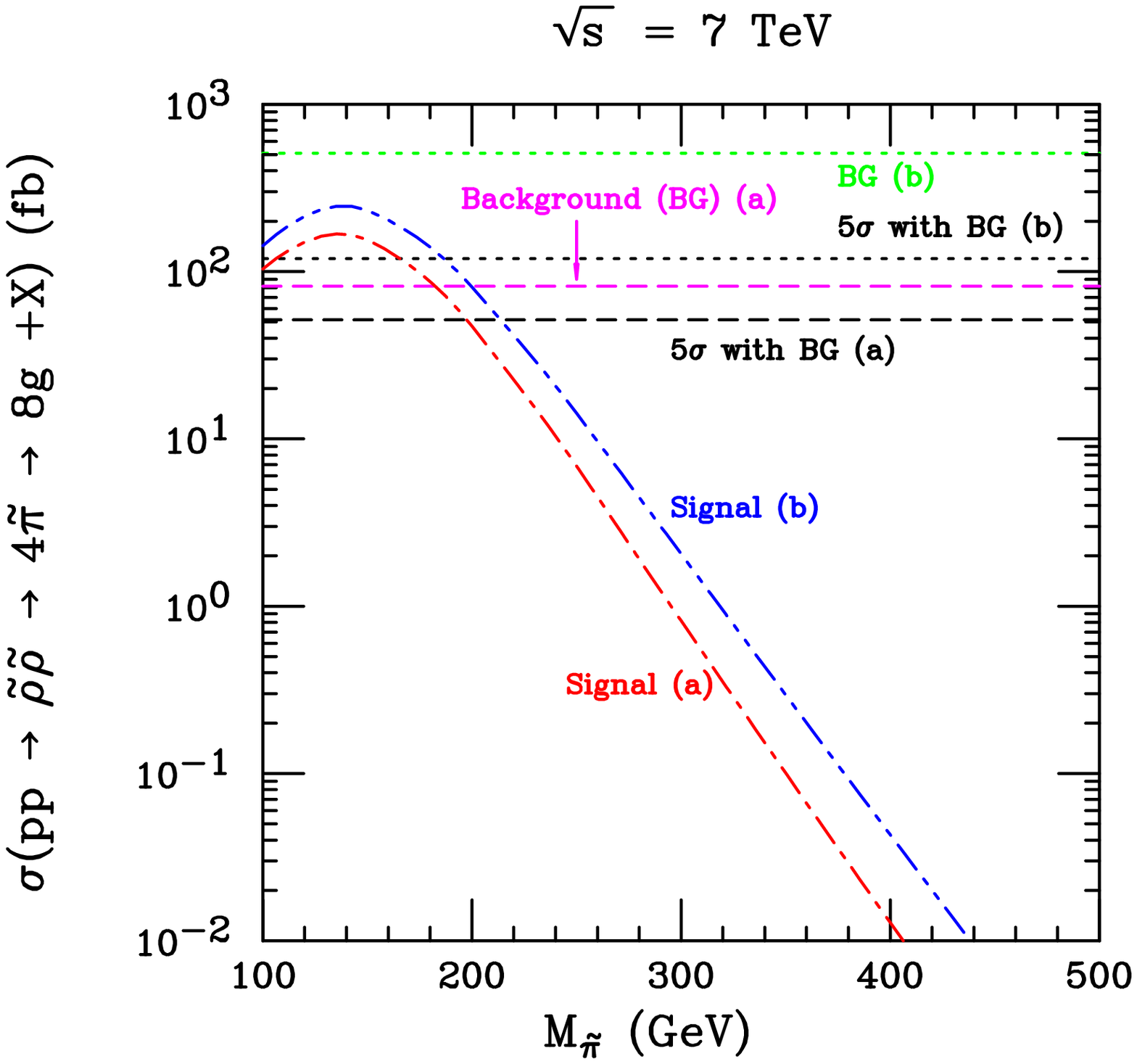}

\caption[]{
The cross section for $pp \to \trho\trho \to 4\tpi \to 8g +X$ 
at the LHC with $\sqrt{s} = 7$ TeV, as a function of $M_{\tpi}$.
We have applied kinematic cuts on $p_T$, $\eta$, $\Del R$ and two sets of
relative mass cuts:
(a) $\Del M_{2j} < 30$ GeV and $\Del M_{4j} < 60$ GeV [red, dot-dash], or
(b) $\Del M_{2j} < 50$ GeV and $\Del M_{4j} < 100$ GeV [blue, dot-dot-dash].
Also shown are the background cross section for the SM processes
from the production of $8g$, $7g1q$, $6g2q$, and $5g3q$ with relative mass
cut (a) [magenta, dash] and relative mass cut (b) [green, dot].
In addition, we present the minimal signal cross section that is required by
a 5 sigma criterion with relative mass cut (a) [dash] and relative mass
cut (b) [dot].
\label{fig:sigma2}
}\end{figure}

Figure 3 shows cross sections of the coloron signal ($\sigma_s$) and
the physics background ($\sigma_b$) from the production of
$8g$, $7g1q$, $6g2q$, and $5g3q$,
with acceptance cuts and fixed mass cuts versus $M_{\tpi}$.
We have replaced the relative mass cuts
of Fig. 2 with two sets of fixed mass cuts:
(a) $|M_{2j} -M_{\tpi}| < 0.10 M_{\tpi}$ and
    $|M_{4j} -M_{\trho}| < 0.15 M_{\trho}$,
or
(b) $|M_{2j} -M_{\tpi}| < 0.15 M_{\tpi}$ and
    $|M_{4j} -M_{\trho}| < 0.20 M_{\trho}$.
Also shown is the minimal signal cross section that is required by
a 5 sigma criterion.
If the background has fewer than 16 events assuming 1 fb$^{-1}$ of
luminosity, we employ the Poisson distribution and require that 
the Poisson probability for the SM background to fluctuate to this
level should be less than $2.85\times 10^{-7}$.
For $M_{\tpi} = 100$ GeV it is very time consuming to get a convergent cross 
section for $\sigma_b$. To improve the stability we have applied 
somewhat less stringent $p_T$ cuts than those given in Eq.~(5),
\begin{equation}
p_T(j_1, \cdots, j_8) \ge 200,150,120,100,80,60,40,20 \; {\rm GeV}
\end{equation}
respectively.
Therefore, the background cross section (red, cross) and the corresponding
$5\sigma$ signal cross section (green, diamond) for $M_{\tpi} = 100$ GeV
are presented with symbols. We note that even with lower $p_T$ cuts,
the background is negligible.

If the ATLAS~\cite{ATLAS} and CMS~\cite{CMS} detectors have excellent
mass resolution, we will be able to apply the narrower fixed mass cut (a),
which has the potential to discover the colorons and hyper-pions up to
$M_{\tpi} = 220$ GeV ($M_{\trho} = 733$ GeV).
A wider fixed mass cut (b) will allow more background events, 
which results in a slightly reduced discovery reach of
$M_{\tpi} = 210$ GeV ($M_{\trho} = 700$ GeV).


\begin{figure}[htb]
\centering\leavevmode
\epsfxsize=4.8in\epsffile{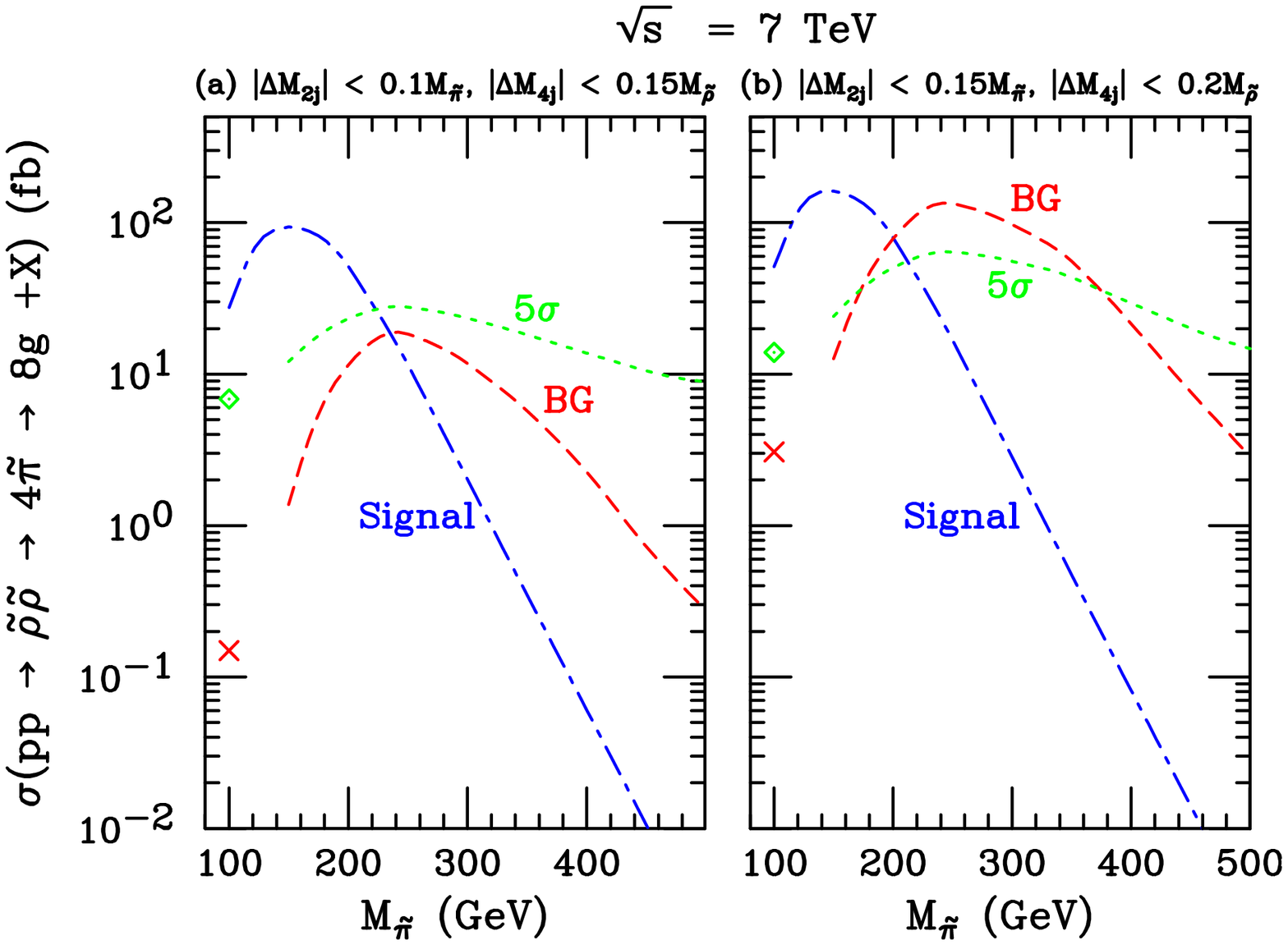}

\caption[]{
The cross section for $pp \to \trho\trho \to 4\tpi \to 8g +X$
(blue, dot-dash) at the LHC with $\sqrt{s} = 7$ TeV,
as a function of $M_{\tpi}$ with acceptance cuts on $p_T$, $\eta$,
and $\Del R$.
We have applied two sets of fixed mass cuts:
(a) $|M_{2j} -M_{\tpi}| < 0.10 M_{\tpi}$ and
    $|M_{4j} -M_{\trho}| < 0.15 M_{\trho}$,
or
(b) $|M_{2j} -M_{\tpi}| < 0.15 M_{\tpi}$ and
    $|M_{4j} -M_{\trho}| < 0.20 M_{\trho}$. 
Also shown are the SM background cross section ($\sigma_b$)(red, dash) and
the minimal signal cross section that is required by a 5 sigma
criterion (green, dot). For $M_{\tpi} = 100$ GeV, the $5\sigma$ signal 
cross section and $\sigma_b$ (green diamond, red cross) 
are calculated with lower $p_T$ cuts.
\label{fig:sigma3}
}\end{figure}
%

\medskip

{\sl Conclusions.}--
We have demonstrated that colorons and hyper-pions can be
produced abundantly at the early stage of the LHC with a center of mass
energy $\sqrt{s} = 7$ TeV and an integrated luminosity of 1 fb$^{-1}$.
With realistic acceptance cuts as well as relative mass cuts
or fixed mass cuts, the physics background can be significantly
reduced to establish a $5\sigma$ signal for
$M_{\tpi} \alt 220$ GeV ($M_{\trho} \alt 733$ GeV).

If the center of mass energy can be raised or the integrated
luminosity increased the discovery potential of colorons at the early LHC
will be significantly improved.
The discovery potential at the LHC for colorons and hyper-pions will
be greatly enhanced with the full center of mass energy $\sqrt{s} = 14$ TeV
and integrated luminosity $L = 30 - 300$ fb$^{-1}$~\cite{Coloron14}.

\medskip

{\sl Acknowledgments.}--
This research was supported
in part by the U.S. Department of Energy
under Grants
No.~DE-FG02-04ER41305,
No.~DE-FG03-93ER40757,
No.~DE-FG02-04ER41306 and
No.~DE-FG02-04ER46140.

\newpage


\end{document}